\documentclass[pdflatex,sn-mathphys-num]{sn-jnl}

\usepackage{graphicx}%
\usepackage{multirow}%
\usepackage{amsmath,amssymb,amsfonts}%
\usepackage{amsthm}%
\usepackage{mathrsfs}%
\usepackage[title]{appendix}%
\usepackage{xcolor}%
\usepackage{textcomp}%
\usepackage{manyfoot}%
\usepackage{svg}%
\usepackage{booktabs}%
\usepackage{algorithm}%
\usepackage{algorithmicx}%
\usepackage{algpseudocode}%
\usepackage{listings}%
\usepackage{subcaption}%
\usepackage[utf8]{inputenc}    

\theoremstyle{thmstyleone}%
\theoremstyle{thmstyletwo}%

\theoremstyle{thmstylethree}%

\begin{document}

\title[Article Title]{Information Abstraction for Data Transmission Networks based on Large Language Models}


\author[1]{\fnm{Haoyuan} \sur{Zhu}}\email{hzhu51@sheffield.ac.uk}

\author*[1]{\fnm{Haonan} \sur{Hu}}\email{haonan.hu@sheffield.ac.uk}

\author*[2,3]{\fnm{Jie} \sur{Zhang}}\email{jie.zhang@ranplanwireless.com}

\affil*[1]{\orgdiv{Department of Electronic and Electrical Engineering}, \orgname{University of Sheffield}, \orgaddress{ \city{Sheffield}, \postcode{S10 2TN}, \country{UK}}}

\affil[2]{\orgdiv{R\&D Department}, \orgname{Cambridge AI+ Ltd.}, \orgaddress{ \city{Cambridge}, \postcode{CB23 3UY}, \country{UK}}}

\affil[3]{\orgdiv{R\&D Department}, \orgname{Ranplan Wireless Network Design Ltd.}, \orgaddress{\city{Cambridge}, \postcode{CB23 3UY} \country{UK}}}


\abstract{Biological systems, particularly the human brain, achieve remarkable energy efficiency by abstracting information across multiple hierarchical levels. In contrast, modern artificial intelligence and communication systems often consume significant energy overheads in transmitting low-level data, with limited emphasis on abstraction. Despite its implicit importance, a formal and computational theory of information abstraction remains absent. In this work, we introduce the Degree of Information Abstraction (DIA), a general metric that quantifies how well a representation compresses input data while preserving task-relevant semantics. We derive a tractable information-theoretic formulation of DIA and propose a DIA-based information abstraction framework. As a case study, we apply DIA to a large language model (LLM)-guided video transmission task, where abstraction-aware encoding significantly reduces transmission volume by $99.75\%$, while maintaining semantic fidelity. Our results suggest that DIA offers a principled tool for rebalancing energy and information in intelligent systems and opens new directions in neural network design, neuromorphic computing, semantic communication, and joint sensing-communication architectures.}

\keywords{Information theory, Semantic communication, Large language model, Neuronal compute}



\maketitle

\section{Introduction}\label{sec1}

The human brain and its neural system represent a pinnacle of energy-efficient biological evolution. Although the brain accounts for only approximately $2\%$ of human body weight, it consumes nearly $20\%$ of the total energy of the body \cite{Attwell2010GlialNeuronalCBF}. Notably, around $95\%$ of this energy supports internal processes, including computation, memory maintenance, synaptic transmission, and neural communication, while less than $5\%$ is devoted to direct sensory signal transmission \cite{howarthUpdatedEnergyBudgets2012}. This allocation suggests that intelligence is not primarily powered by raw sensing or high-fidelity data throughput, but by information abstraction: the ability to extract structured, high-level representations while suppressing irrelevant variability. The 2021 Nobel Prize in Physiology or Medicine, awarded for the discovery of thermosensory receptors \cite{Caterina1997_TRPV1}, further illustrates how biological systems implement layered processing and abstraction—transforming physical stimuli into behaviorally meaningful internal variables, thereby reducing energetic cost while preserving responsiveness.

In contrast, current information and communication technology (ICT) systems exhibit an inverted energy profile. For example, in 5G base stations, the wireless access module may account for up to $82\%$ of total energy consumption, whereas data centres responsible for computation and storage may consume as little as $15\%$ \cite{GSMAIntelligence_EEAB_2024}. More broadly, transmitting one bit over wireless links may require $10^6$ times more energy than computing one bit locally \cite{horowitz11ComputingsEnergy2014a} \cite{huangCloseExaminationPerformance2012}. This mismatch implies that simply scaling modern wireless architectures, where communication dominates the energy budget, will impose severe sustainability limits as networks become increasingly data-hungry and AI-driven.

Motivated by the efficient energy allocation found in biological systems, we contend that a formal theory of information abstraction (IA) is crucial for rethinking energy distribution in next-generation ICT systems. Specifically, employing abstraction to minimize the quantity and precision of transmitted information—without sacrificing task relevance—could shift system energy use away from costly communication to efficient local computation and representation learning.

Over the past decades, information abstraction has been widely explored in machine learning, including representation learning, feature extraction, dimensionality reduction, and hierarchical modeling \cite{Bengio2013Representation}. Despite its central role, a rigorous framework for quantifying the quality of information abstraction remains underdeveloped. Many studies implicitly assess abstraction quality through the level of abstraction—often described along a low-to-high spectrum, where lower levels preserve concrete, detailed properties and higher levels encode more general or conceptual descriptions \cite{EquitzCover1991SuccessiveRefinement,Genewein2015BoundedRationality,ZeilerFergus2014Visualizing}. However, higher abstraction is not necessarily better: an abstraction is only useful if it discards irrelevant variation while retaining information needed for downstream tasks, reasoning, and control. There remains a paucity of works proposing explicit quantitative criteria for abstraction quality. For instance, \cite{kubo2007a} introduced a graph-based metric that quantifies abstraction levels by analysing directional relationships among software patterns. In \cite{cui2006measuring}, histogram-difference and nearest-neighbor measures were proposed to evaluate how statistical distributions and spatial relationships are preserved after abstraction. From an information-theoretic perspective, \cite{tishby1999information} introduced the information bottleneck principle, formalizing abstraction as a trade-off between compressing input and preserving relevant information about a target variable. While influential, the information bottleneck formulation does not by itself yield a universally applicable metric of abstraction quality, and practical deployment often depends on modeling assumptions, e.g., target selection, sufficiency criteria, and tractable variational approximations. Despite these efforts, there is still no established metric for the degree of abstraction of a representation, nor practical tools for guiding model design and system optimization based on abstraction criteria.

More recently, IA has been introduced into ICT through semantic communication, where the objective shifts from faithfully transmitting symbols to conveying task-relevant meaning \cite{xie2021deepsc_tsp,Gunduz2023BeyondBits}. In this view, abstraction is not merely a representation-learning tool but an architectural principle: communication should transmit semantics—compressed representations aligned with user intent, context, and tasks—rather than raw data. While recent developments have been substantial, the current literature still leaves two notable gaps unaddressed. First, existing semantic communication frameworks often rely on end-to-end learned encoders/decoders but provide limited integration with explicit reasoning or world models, making it difficult to support compositional generalization and explainable decision-making \cite{Xie2021DeepSC,Ma2023ExplainableSC,Liang2022ReasoningSC}. Second, the relationship between semantic abstraction and modern generative models (e.g., GANs, diffusion models, and foundation models) is still unclear: these models can generate or reconstruct high-dimensional signals from compact latents, but how such latents should be structured, evaluated, and communicated to support reliable task performance remains largely unexplored \cite{Rombach2022CVPRLDM,Grassucci2023GESCO,Gunduz2023BeyondBits}.

To address these challenges in a high-impact domain, we focus on video transmission, where video traffic has become the dominant driver of network load, placing sustained pressure on bandwidth and motivating the development of more efficient video transmission mechanisms. Sandvine estimates that total global internet traffic is about 33 exabytes per day, with roughly 22 exabytes per day over fixed networks and 11 exabytes per day over mobile networks, while video alone accounts for $39\%$ of fixed downstream volume and $31\%$ of mobile volume \cite{SandvineGIPR2024}. These statistics indicate that even modest efficiency gains in video delivery can translate into material reductions in end-to-end network load. This motivates introducing information abstraction into video transmission so that pixel streams are mapped to task-relevant semantic representations, conveying only the information necessary for the target service and thereby reducing transmitted volume without compromising the semantic utility required by downstream tasks \cite{Gunduz2023BeyondBits,Shao2022TaskOrientedEdgeInference}.

In this work, we introduce a general theoretical framework for evaluating the Degree of Information Abstraction (DIA). Our key contributions are as follows:
1. We formally define the DIA metric and propose a computable formulation that integrates information compression and semantic preservation, while offering two key practical advantages over mutual-information-based objectives: it avoids explicit mutual-information estimation by using tractable divergence terms in a suitable latent space, and it natively supports multimodal inputs through modality-aligned or shared latent space.
2. We demonstrate the utility of this framework in the context of language-model-guided video transmission, where DIA-guided encoding achieves a significant reduction in transmitted data volume while retaining task performance; moreover, we introduce a Video Semantic Differential Stream (VSDS) module that assists the LLM in understanding spatio-temporal semantic information, thereby improving abstraction-aware decision making for transmission.
3. We outline future directions that extend the DIA framework to a range of applications, including: (i) neuromorphic computing, where DIA can inform energy-efficient spiking network design; (ii) semantic communication, enabling abstraction-aware dynamic bit allocation protocols; and (iii) joint sensing and communication, leveraging multimodal abstraction for cross-domain optimization.

The remainder of this article is organized as follows:In Section 2, we define the DIA metric, derive its theoretical underpinnings, and present a neural implementation framework for sequential abstraction based on large language model architectures.
Section 3 evaluates DIA-guided strategies in a case study of vision-language compression.
Section 4 explores potential extensions and cross-domain applications of DIA.
We conclude with a discussion of limitations and future research opportunities.

\section{Theoretical evaluation framework of information abstraction}


In this section, we establish a theoretical evaluation framework of information abstraction.
We first define DIA and examine its consistency with Information Bottleneck (IB) \cite{tishby1999information} optimization under an appropriate choice of latent space, and then highlight key advantages of DIA over IB, including avoiding explicit mutual information estimation and naturally accommodating multi-modal representations.
Building on these insights, we apply DIA to a semantic video transmission task and develop a DIA-guided optimization framework, i.e., Optimization by PROmpting (OPRO) \cite{Yang2024OPRO}, that leverages a large language model to iteratively refine the end-to-end transmission design by directly maximising the DIA objective.
Furthermore, we propose a VSDS module that supplements the main semantic transmission pathway with an explicit residual semantic signal, thereby improving reconstruction fidelity and overall system performance.

\subsection{Degree of Information Abstraction}
\label{sec:dia}

The DIA is composed of two complementary components: a semantic preservation degree, which quantifies how close the semantics of the abstracted data remain to those of the original data, and an information compression rate, which quantifies how much redundancy has been removed by abstraction.

For the information compression rate, we adopt Shannon entropy from information theory to quantify the reduction in uncertainty induced by compression.
Let \(X\) and \(Y\) denote the original data and the corresponding compressed (or abstracted) representation, respectively.
We define the compression rate as
\[
C(X,Y)=1-\frac{H(Y)}{H(X)}.
\]
Here, \(H(\cdot)\) denotes the Shannon entropy, which characterizes the minimum average number of bits required to represent data by measuring its uncertainty.
In practice, we estimate entropy from the empirical symbol distribution: for a discrete variable \(Z\),
\(H(Z)=-\sum_i P_Z(z_i)\log_2 P_Z(z_i)\);
for a continuous variable,
\(H(Z)=-\int_{-\infty}^{\infty} f_Z(z)\log_2 f_Z(z)\mathrm{d}z\),
where \(P_Z(\cdot)\) and \(f_Z(\cdot)\) are the probability mass function (PMF) and probability density function (PDF), respectively \cite{CoverThomas2006EIT}.

A larger \(C\) indicates stronger compression: \(C\to 1\) corresponds to very high compression, whereas \(C\to 0\) indicates little or no compression.

To compare their semantics in a modality-agnostic manner, both $X$ and $Y$ are projected into a shared latent space $S$ by encoders $f_S(\cdot)$ and $g_S(\cdot)$, respectively.
The resulting latent vectors can be interpreted as samples drawn from two empirical distributions on $S$, denoted by $\hat{X}_S$ and $\hat{Y}_S$.
The semantic discrepancy between the original and abstracted data is then measured by the Kullback–Leibler (KL) divergence $D_{\mathrm{KL}}(\hat{X}_S \Vert \hat{Y}_S)$, which quantifies how much information is lost when $\hat{Y}_S$ is used as a surrogate for $\hat{X}_S$ \cite{CoverThomas2006EIT}.
Based on this, we define the semantic preservation degree as
\begin{equation}
\Theta(X, Y, S) = \frac{1}{D_{\mathrm{KL}} \left( \hat{X}_{S} \parallel \hat{Y}_{S} \right)} ,
\label{eq:theta_def}
\end{equation}
so that a smaller statistical discrepancy in the shared latent space corresponds to a larger value of $\Theta(X, Y, S)$ and hence stronger semantic preservation.
In practice, the KL divergence may be regularized by a small positive constant in the denominator to avoid numerical singularities when the two distributions become extremely close.

By combining the semantic preservation degree and the information compression rate, we define the DIA as
\begin{equation}
\Gamma(X, Y, S)
= C(X, Y)\Theta(X, Y, S)
= \Bigl( 1 - \frac{H(Y)}{H(X)} \Bigr)
\frac{1}{D_{\mathrm{KL}}\bigl(\hat{X}_S \Vert \hat{Y}_S\bigr)} .
\label{eq:dia_def}
\end{equation}
A higher value of $\Gamma(X, Y, S)$ indicates that the abstraction achieves a more desirable balance between aggressive compression and faithful semantic preservation.

The design of DIA is closely aligned with the IB principle, which formalizes the notion of compressing an observation while preserving task-relevant information.
In IB, one seeks a representation $T$ of the input $X$ that minimizes the information complexity of $T$ with respect to $X$, while retaining maximal information about a task variable $U$ (e.g., labels, downstream semantics).
A common Lagrangian form of the IB objective is
\begin{equation}
\mathcal{L}_{\mathrm{IB}}(T)
= I(X;T) - \beta I(T;U),
\label{eq:ib_obj}
\end{equation}
where $\beta>0$ controls the compression–relevance trade-off \cite{tishby1999information}.
In our setting, the abstracted data $Y$ plays the role of a bottleneck representation: the compression rate $C(X,Y)$ encourages a reduction of informational complexity, while the semantic preservation degree $\Theta(X,Y,S)$ encourages retention of task-relevant semantics.

Concretely, fix the input $X$ and consider two candidate abstractions $Y_1$ and $Y_2$, with corresponding bottleneck representations $T_i=g_S(Y_i)$ in a shared latent space $S$.

Assume further that the latent space $S$ is chosen such that the KL discrepancy is monotone decreasing in the IB relevance term, namely, there exists a strictly decreasing function $\varphi(\cdot)$ satisfying
\begin{equation}
D_{\mathrm{KL}}\bigl(\hat{X}_S \Vert \hat{Y}_S\bigr)
= \varphi\bigl(I(T;U)\bigr),
\label{eq:kl_mi_mono}
\end{equation}
so that larger $I(T;U)$, which means stronger task-relevant preservation, corresponds to smaller distributional mismatch between $\hat{X}_S$ and $\hat{Y}_S$. Under these conditions, maximizing DIA yields an ordering consistent with minimizing the IB Lagrangian $\mathcal{L}{\mathrm{IB}}(T)=I(X;T)-\beta I(T;U)$: since $\log(\cdot)$ is strictly increasing, maximizing $\Gamma(X,Y,S)$ is equivalent to maximizing
\begin{equation}
\log \Gamma(X,Y,S)
= \log\Bigl(1-\tfrac{H(Y)}{H(X)}\Bigr) - \log D_{\mathrm{KL}}\bigl(\hat{X}_S \Vert \hat{Y}_S\bigr),
\label{eq:log_dia}
\end{equation}
where the first term is monotone in the compression objective (favoring smaller $I(X;Y)$ via smaller $H(Y)$) and, by \eqref{eq:kl_mi_mono}, the second term is strictly increasing in $I(T;U)$. Hence, $\log\Gamma$ is monotone in a combination of a compression term and a relevance term, inducing the same preference direction as the IB objective for any fixed $\beta>0$, and therefore DIA and IB are monotone-order consistent provided that the shared latent space $S$ satisfies the above monotonicity requirement. Moreover, unlike IB, DIA natively accommodates multimodal inputs via the shared latent space $S$ and eliminates the need for explicit mutual-information estimation, thereby avoiding the intractable computation of terms such as $I(T;U)$ in practice.

Practically, the latent space $S$ must be designed such that the divergence computed in $S$ is both semantically meaningful and optimization-stable.
First, since the transmitter and receiver may operate on heterogeneous modalities (e.g., video and text), the embedding geometry must support cross-modal alignment; otherwise, discrepancies in $S$ are dominated by modality-specific domain gaps rather than semantic mismatch.
Second, $S$ must preserve task-relevant semantic factors while suppressing nuisance variations; otherwise, the induced distributions can become artificially close due to semantic collapse, yielding degenerate, trivial alignment despite poor task performance.
Third, the geometry of $S$ must make divergences and distances comparable to semantic mismatch; otherwise, minimizing $D_{\mathrm{KL}}$ may optimize embedding artifacts that do not translate into perceptual or semantic improvements.
Finally, because $D_{\mathrm{KL}}$ is estimated from finite samples, the induced distributions in $S$ must be tractable enough to admit robust and low-variance estimation.
Therefore, the latent space $S$ should satisfy the following principles:
(i) modality invariance: heterogeneous modalities should be mapped into a common geometry;
(ii) semantic sufficiency: $S$ must retain task-relevant factors of variation while discarding nuisance variations;
(iii) comparability: distances/divergences in $S$ should correlate with semantic mismatch (so that KL-based discrepancy is meaningful);
and (iv) distributional tractability: the induced representations should admit stable estimation of $\hat{X}_S$ and $\hat{Y}_S$ (or their parametric approximations) to compute $D_{\mathrm{KL}}$ robustly.

DIA inherits the intuition of IB but offers two practical advantages.
First, DIA is natively multimodal. By projecting $X$ and $Y$ into a shared latent space $S$ with modality-specific encoders, the semantic term remains well-defined across modalities without requiring an explicit joint probabilistic model across heterogeneous data types.
Second, DIA avoids direct computation of mutual information.
In IB, evaluating or estimating $I(T;U)$ typically involves the intractable conditional distribution $p(u|t)$ through
\begin{equation}
I(T;U)=\mathbb{E}_{p(t,u)}\left[\log \frac{p(u|t)}{p(u)}\right],
\label{eq:mi_conditional}
\end{equation}
which is often difficult to model or estimate accurately in high-dimensional settings \cite{Alemi2017DVIB}.
DIA instead uses KL divergence between empirical distributions in $S$ as a semantically aligned and computationally convenient surrogate, thereby avoiding the hardest probabilistic component of classical IB optimization.

\subsection{DIA for semantic video transmission.}
\label{sec:dia_video}
In the semantic video transmission task considered in this work, the input $X$ is a video represented as a temporal sequence of frames, \(X =  \{\mathbf{F}_t \}_{t=1}^{T} \), where \(\mathbf{F}_t \in \mathbb{R}^{H \times W \times C}\), and the abstracted representation $Y$ is a sequence of semantic tokens, latent vectors, or other compressed features produced by the transmitter,
\begin{equation}
Y = \{ \mathbf{y}_t \}_{t=1}^{T}.
\end{equation}
Each frame $\mathbf{I}_t$ and its abstracted counterpart $\mathbf{y}_t$ are projected into the shared latent space $S$ by encoders $f_S$ and $g_S$ to obtain
\begin{equation}
\hat{\mathbf{x}}_t = f_S(\mathbf{F}_t),
\qquad
\hat{\mathbf{y}}_t = g_S(\mathbf{y}_t),
\qquad t = 1, \dots, T.
\end{equation}
According to the DIA definition, the video-level DIA for the semantic video transmission task can be obtained in the following manner. Firstly,
we get the latent vectors $\{ \hat{\mathbf{x}}_t \}_{t=1}^{T}$ and $\{ \hat{\mathbf{y}}_t \}_{t=1}^{T}$ along the temporal dimension, we obtain the empirical latent distributions $\hat{X}_S^{(v)}$ and $\hat{Y}_S^{(v)}$ that summarize the semantics of the original and abstracted video sequences in the chosen latent space.

Secondly, we define the video-level semantic
preservation degree as
\begin{equation}
\Theta_v(X, Y, S)
= \frac{1}{D_{\mathrm{KL}}\bigl(\hat{X}_S^{(v)} \Vert \hat{Y}_S^{(v)}\bigr)} ,
\label{eq:theta_video_def}
\end{equation}
and the corresponding video-level compression rate is defined as
\begin{equation}
C_v(X, Y)
= 1 - \frac{H\bigl(Y^{(v)}\bigr)}{H\bigl(X^{(v)}\bigr)} ,
\label{eq:compression_video_def}
\end{equation}
where $H\bigl(X^{(v)}\bigr)$ and $H\bigl(Y^{(v)}\bigr)$ denote the entropies of the original and abstracted video sequences, respectively, measured either at the bit level or in the latent semantic domain. 

Finally,we obtain the video-level DIA for a transmission instance as
\begin{equation}
\Gamma_v(X, Y, S)
= C_v(X, Y)\Theta_v(X, Y, S)
= \Bigl( 1 - \frac{H\bigl(Y^{(v)}\bigr)}{H\bigl(X^{(v)}\bigr)} \Bigr)
\frac{1}{D_{\mathrm{KL}}\bigl(\hat{X}_S^{(v)} \Vert \hat{Y}_S^{(v)}\bigr)} .
\label{eq:dia_video_def}
\end{equation}
This video-level DIA provides a unified scalar metric for evaluating and optimizing semantic video transmission schemes: it increases when the transmitter successfully reduces the entropy of the video representation while keeping the transmitted sequence's latent-space semantics close to those of the original video.

\subsection{DIA-based OPRO optimization}
\label{sec:dia_opro}

With the definition of the video-level DIA in Section~\ref{sec:dia_video}, we next use it as the optimization objective for an OPRO-based design of the semantic video transmission system.
Let $V$ denote a random video sequence drawn from the source distribution $P_V$, and let $Y(V)$ denote the abstract representation generated by the transmitter under a given system design.
The system design encompasses all adjustable choices in the end-to-end pipeline, including the architecture and hyperparameters of the semantic encoder, the dimensionality of the latent features, quantization and coding parameters, and the allocation of transmission resources across frames and streams.
For notational simplicity, the dependence of $Y(V)$ on these design choices is suppressed.
For a given video $V$, the video-level DIA is expressed as
\begin{equation}
\Gamma_v\bigl(V, Y(V) S\bigr)
= C_v\bigl(V,Y(V)\bigr)
\Theta_v\bigl(V, Y(V) S\bigr),
\label{eq:dia_video_config}
\end{equation}
where $C_v(\cdot)$ and $\Theta_v(\cdot)$ are the compression rate and semantic preservation degree defined in Section~\ref{sec:dia}.

In addition to DIA, the system design determines the amount of transmission resources consumed by the system, including average bit rate, channel usage, and transmit power.
We denote $R(V)$ as the resource consumption associated with transmitting the abstract representation $Y(V)$ for video $V$, where the dependence of $R(V)$ on the system design is also suppressed for clarity.
We impose an average expected transmission traffic constraint
\begin{equation}
\mathbb{E}_{V \sim P_V}\bigl[ R(V) \bigr] \leq R_{\max},
\label{eq:resource_constraint}
\end{equation}
where $R_{\max}$ is a prescribed limit determined by the communication scenario.
The DIA-based system design problem can then be written as

\begin{align}
\max_{Y}\quad 
& J := \mathbb{E}_{V \sim P_V}\!\left[ \Gamma_v\!\left(V, Y(V), S\right) \right]
\label{eq:dia_opro_optimization_obj}\\
\text{s.t.}\quad 
& \mathbb{E}_{V \sim P_V}\!\left[ R(V) \right] \le R_{\max}
\label{eq:dia_opro_optimization_constr}
\end{align}

A critical ingredient in \eqref{eq:dia_video_config}–\eqref{eq:dia_opro_optimization_constr} is the shared latent space $S$, in which the semantic discrepancy term $\Theta_v(\cdot)$ is evaluated in a modality-agnostic manner.

In this work, we instantiate $S$ using CLIP-aligned embeddings.
CLIP adopts a dual-encoder architecture trained with a symmetric contrastive objective, where image/video and text embeddings are $\ell_2$-normalized and compared via temperature-scaled cosine similarity; matched pairs are pulled together while mismatched pairs are pushed apart \cite{Radford2021CLIP}.
This design provides an explicit operational semantics for distances in $S$: similarity scores are the very quantities optimized to discriminate correct from incorrect pairings, making geometric proximity in $S$ a direct proxy for semantic alignment.

This choice is consistent with the latent-space requirements discussed above.
First, modality invariance is promoted because heterogeneous modalities are mapped into a single shared embedding geometry by construction \cite{Radford2021CLIP,Jia2021ALIGN}.
Second, semantic sufficiency is empirically supported by CLIP’s strong zero-shot transfer across diverse downstream tasks and its robustness under distribution shift, indicating that the embedding retains broad, task-relevant semantic factors rather than dataset-specific artifacts \cite{Radford2021CLIP}.
Third, comparability is strengthened by the observation that CLIP embedding similarity has been successfully used as a reference-free metric for caption–image semantic alignment and correlates well with human judgments, suggesting that distances (and hence divergences) in $S$ meaningfully reflect semantic mismatch \cite{Hessel2021CLIPScore}.
Finally, distributional tractability is facilitated by the normalized embedding geometry (bounded norm and cosine-based similarity), which improves numerical conditioning and stabilizes empirical estimation of $\hat{X}_S$ and $\hat{Y}_S$, thereby supporting reliable computation of $D_{\mathrm{KL}}(\hat{X}_S\Vert\hat{Y}_S)$ \cite{Radford2021CLIP}.

More importantly, our DIA–IB consistency analysis assumes an ordinal monotonic correspondence between the KL-based discrepancy in $S$ and the relevance term, as formalized in \eqref{eq:kl_mi_mono}; therefore, it is sufficient that the learned space preserves relative semantic ordering rather than requiring exact probabilistic calibration.
The CLIP ranking paradigm is particularly appropriate in this regard \cite{Radford2021CLIP}, since its contrastive objective is explicitly designed to enforce correct ordering of matched versus mismatched cross-modal pairs, thereby supporting the monotonicity condition in \eqref{eq:kl_mi_mono} and making $\Theta_v(\cdot)$ a reliable surrogate for task-relevant semantic preservation.

Directly solving~\eqref{eq:dia_opro_optimization_obj}–\eqref{eq:dia_opro_optimization_constr} is challenging, because the design space $\mathcal{Z}$ is typically high-dimensional, mixed discrete–continuous, and non-convex, and the DIA objective $\Gamma_v(\cdot)$ is defined implicitly through learned encoders and empirical distributions in the latent space.
To address this difficulty, we adopt the OPRO paradigm, which uses an LLM as a derivative-free, black-box optimizer to iteratively propose improved configurations by conditioning on textual summaries of historical trials and their scalar feedback \cite{Yang2024OPRO}.
The use of OPRO is particularly suitable in our setting because it eliminates the need for task-specific pretraining or fine-tuning of the optimizer: the LLM can be invoked directly via standard API calls to generate candidate configurations, while the DIA score provides an explicit and comparable objective signal for selection and refinement.
This yields a practical advantage for system-level co-design, where the optimization variables are heterogeneous (architecture, hyperparameters, coding, and resource-allocation policies) and the evaluation pipeline is expensive and non-differentiable; OPRO enables rapid iteration with minimal engineering overhead by treating the entire transmitter–receiver chain as a black box and leveraging the general-purpose reasoning capability embedded in modern LLMs.

\begin{figure*}[t]
\centering
\begin{subfigure}[t]{0.49\textwidth}
\centering
\includegraphics[width=\textwidth]{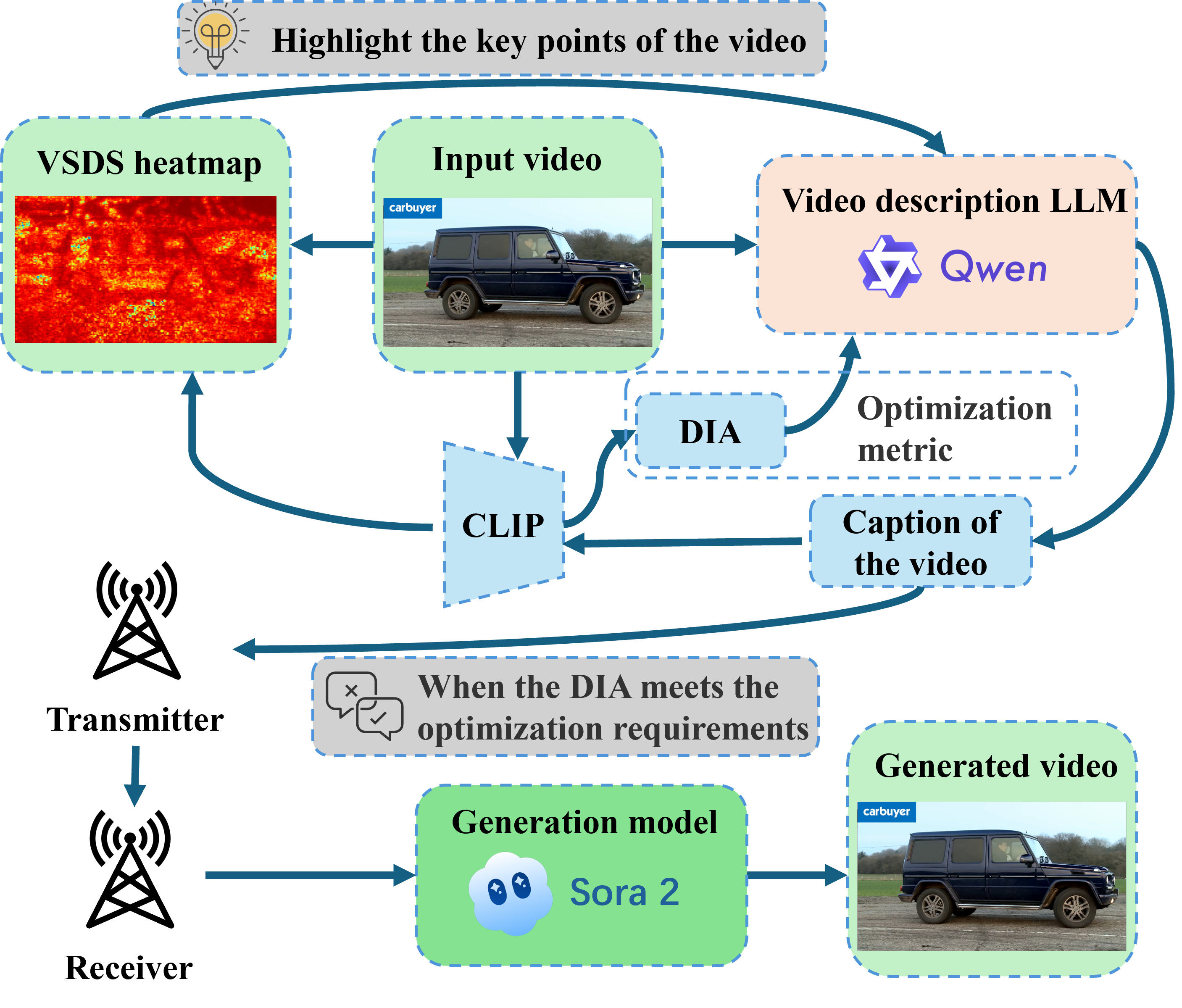}
\caption{DIA-based OPRO optimization workflow.}
\label{fig:dia_opro_workflows:a}
\end{subfigure}
\hfill
\begin{subfigure}[t]{0.49\textwidth}
\centering
\includegraphics[width=\textwidth]{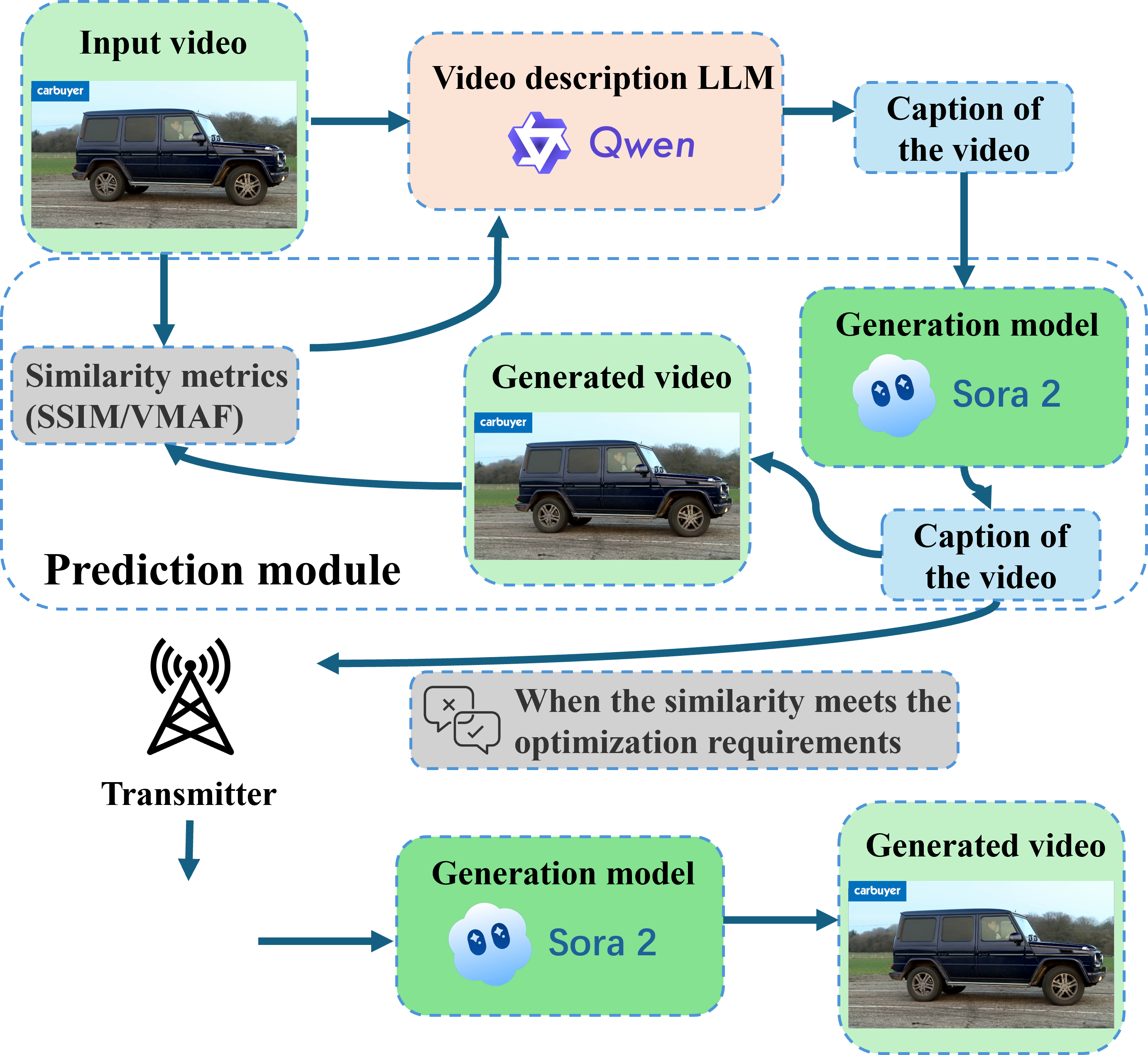}
\caption{Transmitter-side prediction module for caption self-evaluation.}
\label{fig:dia_opro_workflows:b}
\end{subfigure}
\caption{Two transmitter–receiver workflows in DIA-based semantic video transmission.
\subref{fig:dia_opro_workflows:a} depicts the standard pipeline, where DIA provides scalar feedback for OPRO-driven configuration search.
\subref{fig:dia_opro_workflows:b} augments the transmitter with a prediction module that locally generates a video conditioned on the candidate caption to assess its semantic adequacy before transmission, which is preferable when the transmitter has sufficient computation resources.}
\label{fig:dia_opro_workflows}
\end{figure*}

Fig.~\ref{fig:dia_opro_workflows} summarizes two practical workflows for DIA-based OPRO optimization.
In the standard setting (Fig.~\ref{fig:dia_opro_workflows:a}), candidate configurations are instantiated at the transmitter to produce an abstract description, and DIA provides an explicit scalar objective signal for OPRO by quantifying the compression-semantics trade-off under the resource constraint.
When the transmitter side is computation-rich, we further employ a transmitter-side prediction module.  Fig.~\ref{fig:dia_opro_workflows:b} illustrates how the prediction module works. To perform self-evaluation of the candidate caption prior to transmission, the transmitter locally runs a lightweight generation model conditioned on the caption to synthesize a predicted video, and then assesses the semantic adequacy of the caption by comparing the predicted video against the input video.
This predictive loop enables early rejection or refinement of semantically insufficient captions and provides more informative feedback for configuration search, at the cost of additional transmitter-side computation; therefore, it is particularly suitable for scenarios where the transmitter has ample compute budget, whereas the standard pipeline is preferable for compute-limited transmitters.

Specifically, at each OPRO iteration $k$, a set of candidate configurations
\begin{equation}
\mathcal{Z}^{(k)} = \{ \mathbf{z}^{(k,1)}, \dots, \mathbf{z}^{(k,M)} \}
\end{equation}
is evaluated on a batch of training videos, yielding DIA scores
\begin{equation}
J\bigl(\mathbf{z}^{(k,m)}\bigr), \quad m = 1, \dots, M.
\end{equation}
The historical configurations and their scores are then summarized into a textual prompt, which is fed into the LLM together with a natural-language description of the design constraints and performance goals.
Conditioned on this prompt, the LLM generates a new set of candidate configurations
\begin{equation}
\mathcal{Z}^{(k+1)} = \{ \mathbf{z}^{(k+1,1)}, \dots, \mathbf{z}^{(k+1,M)} \},
\end{equation}
which are expected to yield higher DIA values under the resource constraint~\eqref{eq:resource_constraint}.
This process is repeated until the improvement in $J(\mathbf{z})$ saturates or a predefined iteration budget is reached.
In this way, the OPRO algorithm uses DIA as an explicit scalar objective and leverages the reasoning ability of the LLM to explore promising regions of $\mathcal{Z}$, thereby obtaining a configuration $\mathbf{z}^{\star}$ that realizes a favorable trade-off between compression and semantic fidelity for video transmission.

\begin{algorithm}[t]
\caption{DIA-based OPRO Optimization (with Optional Transmitter-side Prediction)}
\label{alg:dia_opro}
\begin{algorithmic}[1]
\Require Training videos (or sampler) $P_V$, design space $\mathcal{Z}$, latent space $S$, budget $R_{\max}$, population size $M$, iterations $K$, batch size $B$, prediction flag $\textsc{Pred}\in{0,1}$, transmitter-side generator $\mathcal{G}(\cdot)$ (if $\textsc{Pred}=1$).
\Ensure Best configuration $\mathbf{z}^{\star}$.
\State Initialize history buffer $\mathcal{H}\gets\emptyset$; initialize $\mathcal{Z}^{(0)}=\{\mathbf{z}^{(0,m)}\}_{m=1}^M$.
\State $J^{\star}\gets -\infty$, $\mathbf{z}^{\star}\gets \texttt{None}$.
\For{$k=0$ to $K-1$}
\For{$m=1$ to $M$}
\State Sample a batch $\mathcal{B}\sim P_V$ with $|\mathcal{B}|=B$.
\State Initialize accumulators: $\hat{J}\gets 0$, $\hat{R}\gets 0$.
\ForAll{$V\in\mathcal{B}$}
\State Generate abstraction $Y_{\mathbf{z}^{(k,m)}}(V)$ at the transmitter.
\If{$\textsc{Pred}=1$}
\State Locally synthesize a predicted video $\tilde{V}\gets \mathcal{G}\bigl(Y_{\mathbf{z}^{(k,m)}}(V)\bigr)$.
\State Compute $\Gamma_v \gets \Gamma_v\bigl(V, Y_{\mathbf{z}^{(k,m)}}(V), S; \tilde{V}\bigr)$ \Comment{DIA evaluated with prediction-assisted semantic assessment}
\Else
\State Compute $\Gamma_v \gets \Gamma_v\bigl(V, Y_{\mathbf{z}^{(k,m)}}(V), S\bigr)$.
\EndIf
\State $\hat{J}\gets \hat{J}+\Gamma_v$.
\State $\hat{R}\gets \hat{R}+R_{\mathbf{z}^{(k,m)}}(V)$.
\EndFor
\State $J(\mathbf{z}^{(k,m)}) \gets \hat{J}/B$, \quad $\bar{R}(\mathbf{z}^{(k,m)}) \gets \hat{R}/B$.
\State Append $(\mathbf{z}^{(k,m)},J(\mathbf{z}^{(k,m)}),\bar{R}(\mathbf{z}^{(k,m)}))$ to $\mathcal{H}$.
\If{$\bar{R}(\mathbf{z}^{(k,m)})\le R_{\max}$ \textbf{and} $J(\mathbf{z}^{(k,m)})>J^{\star}$}
\State $\mathbf{z}^{\star}\gets \mathbf{z}^{(k,m)}$, \quad $J^{\star}\gets J(\mathbf{z}^{(k,m)})$.
\EndIf
\EndFor
\State Build a prompt $\mathcal{P}^{(k)}$ from $\mathcal{H}$, including top candidates, their DIA scores, and constraint violations.
\State Query the LLM via API with $\mathcal{P}^{(k)}$ to obtain $\mathcal{Z}^{(k+1)}=\{\mathbf{z}^{(k+1,m)}\}_{m=1}^M$.
\EndFor
\State \Return $\mathbf{z}^{\star}$.
\end{algorithmic}
\end{algorithm}

\subsection{Evaluation metrics}
We evaluate the similarity between the reconstructed and source videos using two complementary metrics.
First, we report the structural similarity index measure (SSIM), which compares local luminance, contrast, and structural statistics between two signals and is commonly used as a reference-based indicator of frame-level structural fidelity \cite{wang2004ssim}.
Second, we report VMAF, a full-reference video quality metric derived from spatio-temporal perceptual feature modeling, which is designed to capture video-level perceptual quality and semantic consistency beyond purely pixel-wise distortions \cite{bampis2018st_vmaf,mittag2023lstm_vqa}.
We choose SSIM and VMAF because they provide complementary views of reconstruction quality: SSIM emphasizes low-level structural preservation, whereas VMAF is more sensitive to perceptual and temporally aggregated distortions that are particularly relevant for generative or semantic reconstruction \cite{bampis2018st_vmaf, mittag2023lstm_vqa, wang2004ssim}.

\subsection{Video semantic differential stream (VSDS) module}
\label{sec:semantic_diff_stream}

To provide more temporally coherent semantic information for both the DIA-based OPRO optimization in Section~\ref{sec:dia_opro} and the IB-based LLM-driven framework, we introduce a semantic differential stream module.
The goal is two-fold: first, to enhance the temporal continuity of the representation by explicitly modeling how semantics evolve over time; and second, to guide the LLM during description generation to focus more on regions with strong motion, thereby enriching the resulting captions with motion-related details.
This module illustrates how semantic representations of the video evolve over time and relates these changes to pixel-level temporal differences.

Consider a video sequence $V = \{ \mathbf{F}_t \}_{t=1}^{T}$, where $\mathbf{F}_t \in \mathbb{R}^{H \times W \times C}$
and let $\phi(\cdot)$ denote a pre-trained CLIP-based visual encoder that maps each frame into a $d$-dimensional semantic embedding:
\begin{equation}
\mathbf{v}_t = \phi(\mathbf{F}_t) \in \mathbb{R}^{d},
\qquad t = 1, \dots, T.
\end{equation}
To characterize how the global semantics of the video evolve over time, we approximate the temporal derivative of the semantic representation using finite differences,
\begin{equation}
\mathbf{A}_t
= \mathbf{v}_t - \mathbf{v}_{t-1},
\qquad t = 2, \dots, T,
\label{eq:semantic_derivative}
\end{equation}
where $\mathbf{A}_t$ measures the frame-to-frame semantic change in the CLIP embedding space.

In parallel, we quantify temporal changes in the pixel domain.
Each frame $\mathbf{F}_t$ is vectorized into
\begin{equation}
\mathbf{p}_t = \mathrm{vec}(\mathbf{F}_t) \in \mathbb{R}^{N},
\qquad N = H \times W \times C,
\end{equation}
and the corresponding pixel-level temporal difference is defined as
\begin{equation}
\mathbf{B}_t
= \mathbf{p}_t - \mathbf{p}_{t-1},
\qquad t = 2, \dots, T.
\label{eq:pixel_derivative}
\end{equation}
We assume a locally linear relationship between pixel-level changes and semantic changes,
\begin{equation}
\mathbf{A}_t = \mathbf{C}_t \mathbf{B}_t,
\qquad t = 2, \dots, T,
\label{eq:linear_relation}
\end{equation}
where $\mathbf{C}_t \in \mathbb{R}^{d \times N}$ is a time-varying sensitivity matrix, where $d$ denotes the dimensionality of the semantic embedding space and $N$ denotes the number of elements of the pixel-domain (e.g., pixels or vectorized spatio-temporal samples) whose temporal differences are mapped, which maps temporal differences from the pixel-domain to temporal differences from the semantic-domain.

Given the observed pair $\bigl(\mathbf{A}_t, \mathbf{B}_t\bigr)$, we estimate $\mathbf{C}_t$ via regularized least squares,
\begin{equation}
\mathbf{C}_t
= \arg\min_{\mathbf{C}}
\bigl|\bigl| \mathbf{A}_t - \mathbf{C}\mathbf{B}_t \bigr|\bigl|_2^2
+ \lambda \bigl|\bigl| \mathbf{C} \bigl|\bigr|_F^2,
\label{eq:C_estimation}
\end{equation}
with regularization parameter $\lambda>0$.

The regularization term $\lambda||\mathbf{C}||_F^2$ is introduced to mitigate the ill-posedness and numerical instability of estimating $\mathbf{C}_t\in\mathbb{R}^{d\times N}$ from limited observations. In our setting, $N$ (the pixel-domain dimension) is typically orders of magnitude larger than $d$, and at each time $t$ we only observe a single pair $(\mathbf{A}_t,\mathbf{B}_t)$, so the unregularized least-squares problem is severely underdetermined and admits infinitely many solutions. As a consequence, small perturbations or noise in $\mathbf{A}_t$ and $\mathbf{B}_t$ can lead to arbitrarily significant changes in the fitted $\mathbf{C}_t$, resulting in overfitting and unstable sensitivity estimates. The $\ell_2$ (ridge/Tikhonov) penalty yields a unique minimum-norm solution. It improves conditioning by shrinking the magnitude of $\mathbf{C}_t$, thereby stabilizing the mapping from pixel-level temporal differences to semantic-level differences and producing more robust, interpretable saliency patterns. The hyperparameter $\lambda>0$ controls the bias–variance trade-off and can be selected via standard criteria such as cross-validation \cite{HoerlKennard1970Ridge,TikhonovArsenin1977IllPosed,GolubHeathWahba1979GCV}
.

The semantic importance of each pixel (or spatial location) at time $t$ is derived from the columns of $\mathbf{C}_t$.
Let $\mathbf{c}_{t,n} \in \mathbb{R}^{d}$ denote the $n$-th column of $\mathbf{C}_t$.
We define a scalar sensitivity score
\begin{equation}
s_{t,n} = \bigl|\bigl| \mathbf{c}_{t,n} \bigl|\bigr|_2,
\qquad n = 1, \dots, N,
\label{eq:sensitivity_score}
\end{equation}
and collect these scores into $\mathbf{s}_t \in \mathbb{R}^{N}$.
By reshaping $\mathbf{s}_t$ back to the spatial dimensions $(H, W)$ and optionally aggregating over channels, we obtain a semantic heatmap
\begin{equation}
\mathbf{H}_t \in \mathbb{R}^{H \times W},
\qquad t = 2, \dots, T,
\label{eq:heatmap_def}
\end{equation}
which highlights regions whose temporal variations exert the strongest influence on the CLIP semantics.
The resulting sequence of heatmaps
\begin{equation}
\mathcal{D}(V)
= \{ \mathbf{H}_t \}_{t=2}^{T}
\end{equation}
is referred to as the semantic differential stream associated with video $V$.

The semantic differential stream can be integrated into both optimization frameworks via an explicit VSDS-guided region-prioritization pipeline.
Concretely, we first compute the VSDS heatmaps $\{\mathbf{H}_t\}_{t=2}^{T}$ defined in \eqref{eq:heatmap_def}–\eqref{eq:heatmap_def}.
We then partition the original video into a set of non-overlapping spatio-temporal blocks $\{\mathcal{R}_k\}_{k=1}^{K}$ (e.g., a uniform spatial grid for each frame, optionally grouped into short temporal windows), and assign each block an importance score by aggregating the VSDS intensity over its support, for instance
$w_k = \sum_{t}\sum_{(i,j)\in \mathcal{R}_k}\mathbf{H}_t(i,j)$.
Sorting ${w_k}$ yields an importance ordering $\pi$ over blocks, where earlier elements correspond to regions with stronger semantic dynamics.
In the DIA-guided OPRO setting, this ordering provides a direct spatio-temporal saliency signal for allocating representation capacity and transmission resources preferentially to high-ranked blocks, improving temporal continuity and reducing semantic drift across frames.
In the IB-driven LLM-based method, we inject the same ordering $\pi$ into the LLM prompt (together with the corresponding block indices and coordinates), instructing the description model to attend to and describe the high-ranked blocks first; this explicitly steers caption generation toward motion-dominant, semantically changing regions and therefore yields more motion-aware descriptions for reconstruction.
In both cases, VSDS operationalizes the temporal evolution of semantics into an actionable prioritization mechanism, improving continuity and motion fidelity in the reconstructed video.

\section{Experiments}
\label{sec:experiments}

This section evaluates the DIA-driven semantic video transmission framework under a unified five-round OPRO schedule on a fixed multi-category video set. We compare four strategies (local baseline, DIA-OPRO, an IB-based counterpart, and VSDS-augmented DIA-OPRO) and report SSIM, VMAF, and DIA to quantify fidelity, perceptual/semantic alignment, and abstraction quality.

\subsection{Experimental Setup}
\label{subsec:exp_setup}

\subsubsection{Model instantiation and system components}

We consider a semantic video transmission task where the transmitter extracts and conveys compact semantic information from a source video, and the receiver reconstructs a video that preserves both structural fidelity and semantic consistency with the source.
Following the proposed design, we adopt the video-level DIA as the primary optimization objective and use it to drive an OPRO-style design search over system configurations.
Concretely, at each optimization round, OPRO proposes candidate configuration updates and selects the candidate that maximizes the DIA score aggregated over the evaluation set.

The video understanding and semantic extraction module at the transmitter is implemented using Qianwen Qwen-VL-Max \cite{Bai2023QwenVL}.
It is responsible for producing high-level semantic abstractions that condition downstream semantic transmission and receiver-side reconstruction.

Both the receiver-side video generation model and the prediction module used in the closed-loop evaluation are implemented with sora-video2-landscape \cite{OpenAI2025Sora2Release}.
The generated video resolution is fixed to $1280\times704$ pixels, ensuring that all evaluated reconstructions are produced under a consistent rendering distribution and avoiding confounding effects introduced by heterogeneous generative backends.

The latent space used in DIA and the semantic embedding space used in the VSDS module are both implemented with CLIP RN-50 \cite{Radford2021CLIP}.
Using the same semantic space for DIA and the differential semantic stream ensures metric-space alignment between the optimization signal and the auxiliary semantic-correction pathway.
The reason why CLIP is appropriate for DIA is illustrated in \ref{sec:dia_opro}.

Since the receiver generates videos at a fixed resolution of 1280$\times$704 pixels, we construct an evaluation dataset consisting of 20 videos at the same resolution.
To reduce content bias and to probe robustness across distinct semantic structures, the dataset contains two coarse-grained categories:
(i) videos dominated by human subjects and actions, and
(ii) videos dominated by objects and scene dynamics.
We choose these two categories to improve dataset coverage and to ensure that evaluation reflects complementary sources of semantic variation: human-centric videos emphasize articulated motion, fine-grained actions, and agent–object interactions, whereas object/scene-centric videos emphasize background dynamics, camera motion, and changes in scene composition.
By jointly covering these two regimes, the evaluation better reflects the diversity of real-world video semantics and reduces the risk that conclusions are driven by a single content type.
All methods are evaluated on the same set of 20 videos at every optimization round.

\subsubsection{Compared methods and optimization schedule}
We compare four optimization strategies:
\begin{itemize}
\item \textbf{Prediction module optimization}: a local baseline that optimizes only the prediction module. Concretely, the prediction module takes the transmitter's semantic abstraction and generates a set of candidate textual descriptions to condition the receiver-side video generator. For each candidate description, the receiver synthesizes a video, and the module evaluates the reconstruction by computing SSIM and VMAF scores between the generated and source videos. The candidate that maximizes the chosen evaluation objective, implemented as an SSIM/VMAF-based score, is then selected to update the prompt-level description for subsequent rounds.

\item \textbf{OPRO(DIA)}: OPRO-driven optimization using DIA as the objective.

\item \textbf{IB}: an optimization strategy based on an IB architecture, used as a theoretical and structural counterpart to validate consistency with DIA-driven optimization.

\item \textbf{VSDS-OPRO}: OPRO-driven optimization augmented with the VSDS.
\end{itemize}

These four strategies are selected to provide controlled, hypothesis-driven comparisons. The optimization of the prediction module serves as a reference baseline that isolates gains attainable by local prompt selection without a global configuration search. Comparing OPRO (DIA) with its IB-based counterpart provides an explicit empirical test of the claimed consistency between DIA- and IB-driven optimization when the latent space is shared. Finally, contrasting VSDS-OPRO with standard OPRO(DIA) directly tests the effectiveness of the VSDS module, i.e., whether incorporating differential spatio-temporal semantic cues yields additional improvements beyond the main OPRO updates.

All methods are run for \textbf{5} optimization rounds, and the metrics are recorded after each round.

\subsection{Experimental Results}
\label{subsec:exp_results}

\begin{figure}[t]
\centering
\includegraphics[width=0.72\linewidth]{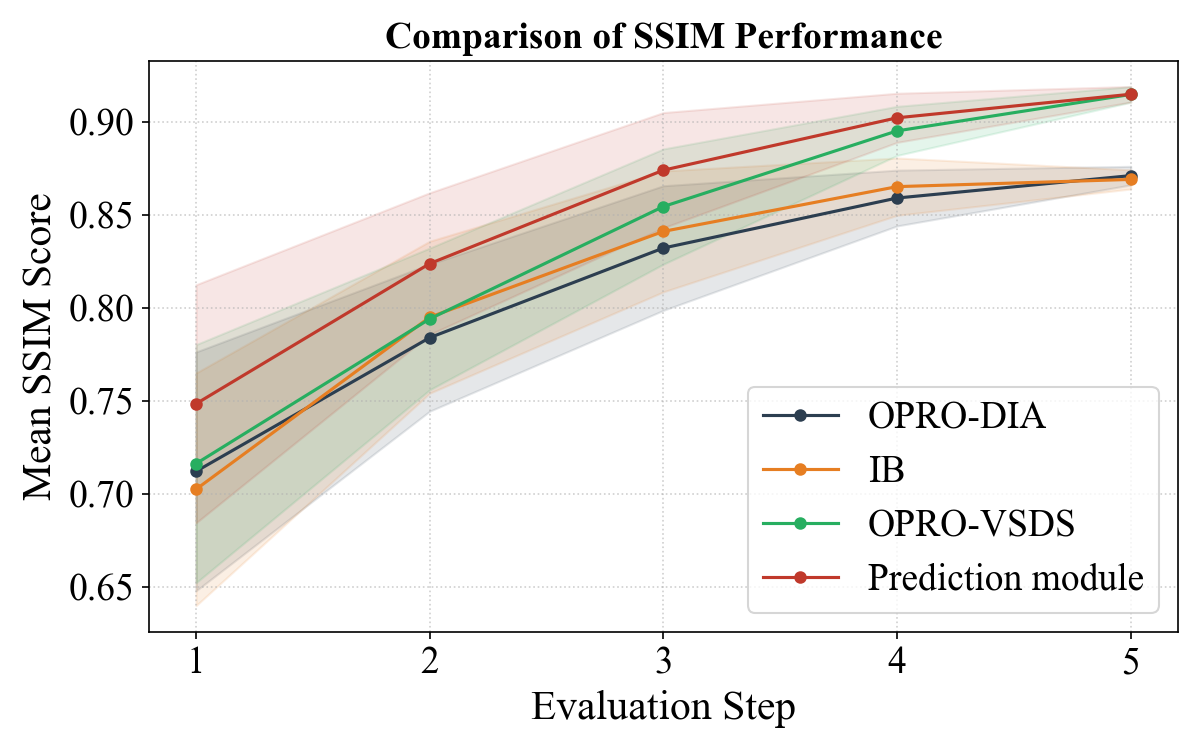}
\caption{SSIM versus optimization round for different optimization strategies.}
\label{fig:ssim_round}
\end{figure}

Fig.~\ref{fig:ssim_round} shows SSIM versus optimization round. All methods maintain high SSIM values and converge within a narrow range, indicating that the reconstructed videos preserve structural fidelity well throughout optimization. This convergence and cross-method consistency further suggest that using DIA as the optimization objective in OPRO yields strong performance on structure-aware full-reference metrics. In addition, OPRO with VSDS achieves slightly higher SSIM than vanilla OPRO in later rounds, supporting the effectiveness of VSDS in improving the structural quality of regenerated videos beyond the gains from OPRO updates alone. The narrow inter-method gap is expected because SSIM mainly reflects low-level structural similarity, while our receiver uses a fixed generative prior at a fixed resolution that already preserves global geometry well; therefore, most optimization gains manifest in motion/semantic details rather than in coarse structure.

\begin{figure}[t]
\centering
\includegraphics[width=0.72\linewidth]{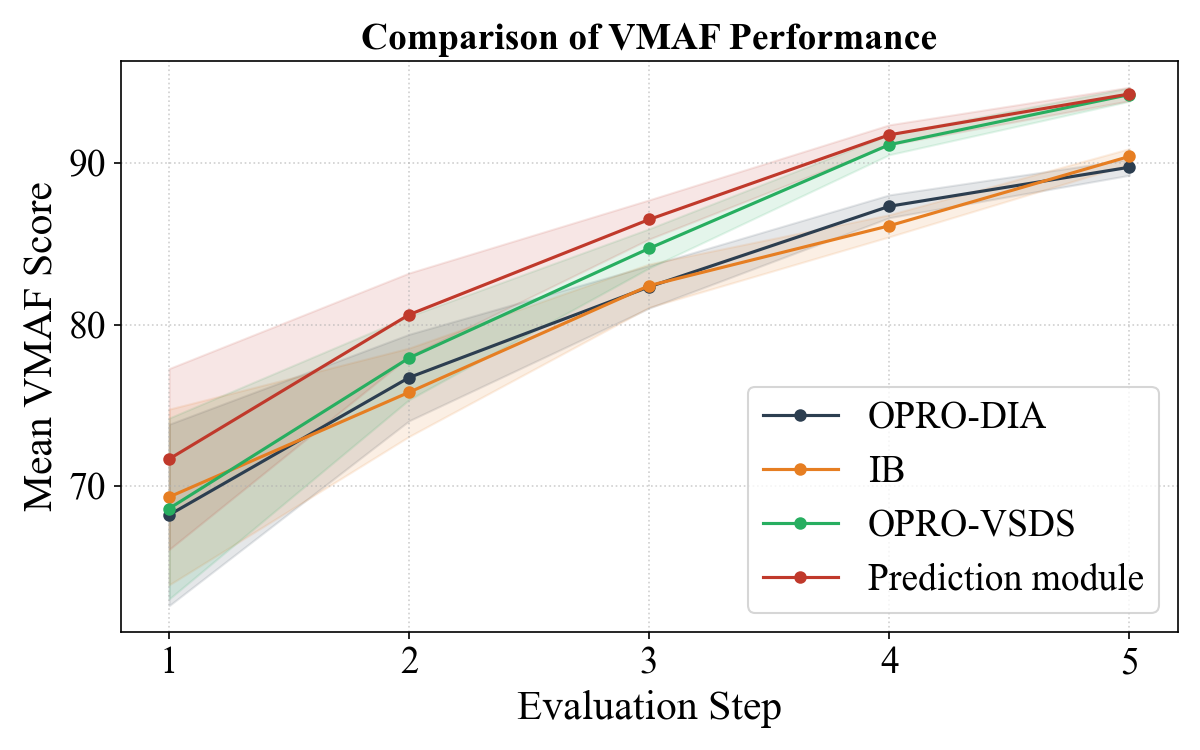}
\caption{VMAF versus optimization round for different optimization strategies.}
\label{fig:VMAF_round}
\end{figure}

Fig.~\ref{fig:VMAF_round} reports VMAF versus optimization round. Similar to SSIM, VMAF exhibits a consistent increasing trend and converges in the final rounds across all methods, indicating that the optimization process improves video-level perceptual and semantic alignment as captured by a full-reference metric. The convergence and consistency of VMAF imply that DIA-guided OPRO performs well under video-level quality evaluation. Moreover, OPRO with VSDS is marginally higher than vanilla OPRO, providing evidence that VSDS also improves full-reference video quality, rather than only low-level structural fidelity. The early-round improvement suggests that OPRO quickly corrects missing or ambiguous semantic factors in the conditioning descriptions, whereas the late-round plateau indicates diminishing returns due to the generator's capacity and the limited abstraction budget.

\begin{figure}[t]
\centering
\includegraphics[width=0.72\linewidth]{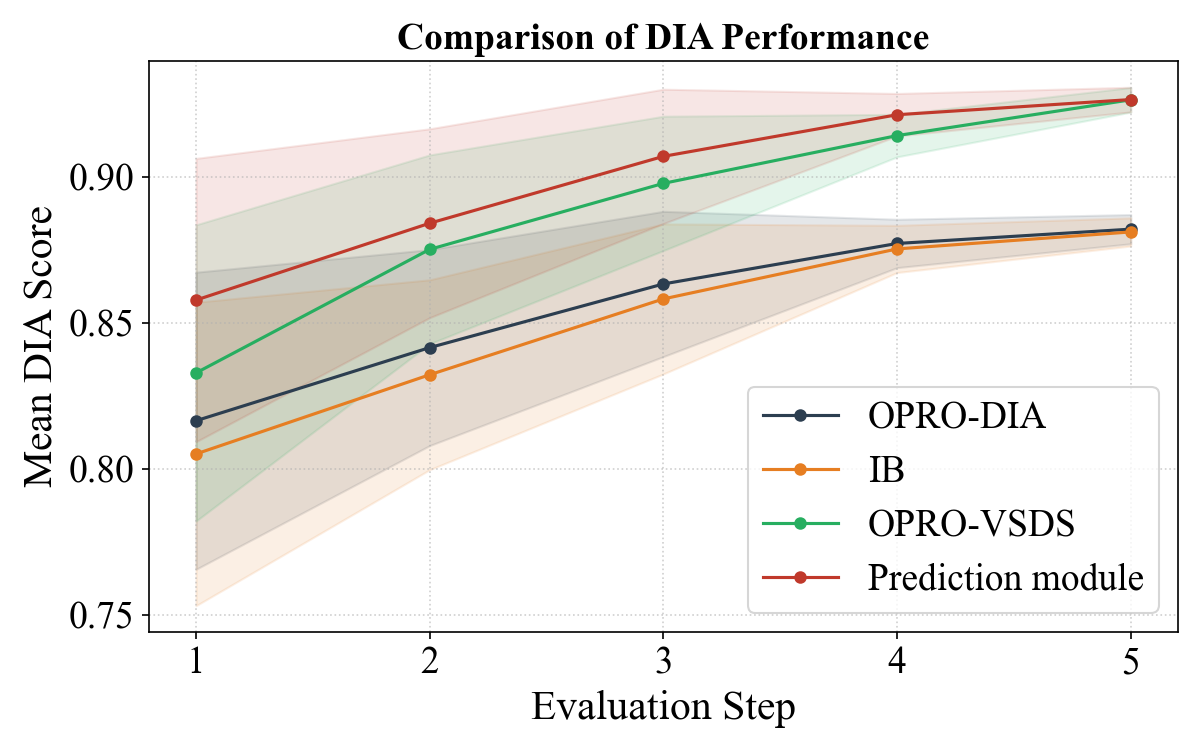}
\caption{DIA versus optimization round for different optimization strategies.}
\label{fig:dia_round}
\end{figure}

Fig.~\ref{fig:dia_round} presents DIA versus optimization round. Both OPRO and VSDS-OPRO exhibit monotonic improvement and clear convergence, and their DIA trajectories are consistent with the improvements and stabilization observed in SSIM and VMAF. This agreement indicates that DIA is a reasonable and effective optimization target for semantic video transmission in this experimental instantiation. Furthermore, VSDS-OPRO achieves the highest DIA values in mid-to-late rounds, demonstrating that the VSDS module directly enhances abstraction quality by capturing and compensating for spatio-temporal semantic residuals beyond the primary OPRO updates.

Overall, the consistent monotonic improvements of OPRO and VSDS-OPRO relative to the prediction-module baseline across SSIM, VMAF, and DIA support the practicality and effectiveness of DIA-driven optimization. At the same time, the close alignment between the trajectories of OPRO/VSDS-OPRO and the IB-based strategy across all three metrics provides experimental evidence that optimizing DIA induces optimization comparable to that of an IB-based objective under the shared latent space and end-to-end system, thereby validating the consistency between IB and DIA in semantic video transmission. The advantage becomes more visible in mid-to-late rounds because once global semantics are largely resolved, the remaining errors concentrate on motion-dominant local regions; VSDS explicitly prioritizes these spatio-temporal residuals and hence provides additional, albeit marginal, gains. Moreover, the 95\% confidence intervals of SSIM, VMAF, and DIA consistently contract as the optimization proceeds. Since the evaluation set is fixed across rounds, this narrowing indicates a reduction in per-round performance dispersion (variance) under each optimization strategy, implying that the learned configurations become progressively less sensitive to content-dependent fluctuations. This behavior provides quantitative evidence of optimization stability and convergence, complementing the observed improvements in the mean trajectories.

\section{Discussion}

The DIA framework offers a novel theoretical lens for quantifying and managing the trade-off between semantic fidelity and resource efficiency in intelligent systems. By introducing a task- and data-aware metric of abstraction, this framework reorients the design of artificial neural networks, communication protocols, and multimodal perception systems toward co-optimization of abstraction efficiency. Our results demonstrate that, when guided by DIA, image transmission systems can achieve up to $99.75\%$ reduction in transmitted bits without perceptual loss, as judged by downstream task performance, underscoring DIA’s potential as a unifying principle across disciplines.

In neural network design, DIA can serve as an information-theoretic compass for architectural choices. As a widely adopted Transformer backbone for language understanding, Bidirectional Encoder Representations from Transformers (BERT) provides a natural testbed for examining DIA-guided architectural optimization. Building on the BERT architecture \cite{devlin-etal-2019-bert}, we anticipate that DIA profiling can guide hierarchical layer pruning, attention-head reallocation, and principled bottleneck sizing; developing and validating these DIA-driven compression and reconfiguration strategies will constitute a key focus of our next-stage work. Importantly, this methodology generalizes beyond Transformers. In spiking neural networks (SNNs), which inherently operate under energy constraints \cite{Roy2019_Nature_SNN}, DIA can guide the allocation of spatio-temporal resolution and spike sparsity. For instance, by aligning the abstraction level of spike trains with task-specific DIA thresholds, neuromorphic systems could preserve core semantic content with minimal event-level redundancy—an essential step toward bio-inspired learning and inference.

The abstraction-oriented perspective also challenges the traditional boundary between communication and cognition. Semantic communication, which aims to transmit meaning rather than raw data \cite{gaoOpticalSemanticCommunication2025}, is a natural application of the DIA framework. Unlike conventional entropy-based metrics, DIA is sensitive to both representational hierarchy and semantic utility, enabling dynamic bit allocation strategies that reflect the abstraction levels of transmitted content. This provides a principled basis for tasks such as low-bandwidth language grounding or real-time dialogue systems, where abstraction-adjusted compression directly translates to communicative efficiency.

DIA further enables integrated design across sensing and communication systems, particularly in ISAC paradigms \cite{liuIntegratedSensingCommunications2022}. Multimodal perception tasks—such as autonomous driving or AR/VR navigation—require compressing and fusing heterogeneous data streams while preserving semantic coherence \cite{piechockiMultimodalSensorFusion2023}. Here, abstraction fusion across modalities (e.g., visual, auditory, radar) becomes essential \cite{tanBioinspiredMultisensoryNeural2021}. The DIA framework can guide the selection of shared latent spaces where abstraction levels are matched or harmonized, enabling joint encoding strategies that maximize both interpretability and transmission efficiency.

In the context of image and video transmission, the DIA framework challenges the dominance of pixel-level fidelity as the primary optimization objective. Instead, it introduces a content-aware strategy that privileges abstraction-preserving compression. We anticipate that future codecs and neural image compression models will incorporate DIA as either a regularization term or an explicit loss component to maintain perceptual quality at drastically lower bitrates. Furthermore, coupling DIA with large generative models (e.g., diffusion transformers or semantic autoencoders) can enable reconstruction of abstracted inputs at the receiver, closing the loop between lossy compression and semantic restoration \cite{wangSemanticAwareAutoEncodersSelfsupervised2022a}.

Looking ahead, the DIA framework opens multiple avenues for research. First, the development of differentiable DIA estimators would facilitate end-to-end training of abstraction-aware models. Second, integrating DIA into neural architecture search (NAS) and meta-learning pipelines could yield architectures that self-adapt to varying abstraction demands \cite{elskenNeuralArchitectureSearch2019}\cite{hospedalesMetaLearningNeuralNetworks2022}. Finally, from a cognitive modelling perspective, DIA provides a pathway to formalize aspects of hierarchical abstraction in the brain, offering a quantitative substrate for aligning artificial networks with biological computation.

By re-entering the role of abstraction in information processing, the DIA framework provides a cross-domain principle for designing more intelligent, efficient, and communicative systems. It bridges theory and practice, inviting collaboration across machine learning, signal processing, and cognitive neuroscience.

\bibliography{sn-bibliography}

\end{document}